# Monte Carlo simulations and critical behavior of the Rare Earth Based Alloy GdN

**S. IDRISSI[1*], S. ZITI[2], H. LABRIM[3] and L. BAHMAD[1,*]**

[1] Laboratoire de la Matière Condensée et des Sciences Interdisciplinaires (LaMCScI), Mohammed V University of Rabat, Faculty of Sciences, B.P. 1014 Rabat, Morocco.

[2] Intelligent Processing and Security of Systems, Mohammed V University of Rabat, Faculty of Sciences, B.P. 1014 Rabat, Morocco.

[3] USM/DERS/ Centre National de l'Energie, des Sciences et des Techniques Nucléaires (CNESTEN), Rabat, Morocco.

**Abstract:**

Using Monte Carlo Simulations (MCS), based the Metropolis algorithm, we examine in this study, the magnetic properties of the rare earth based compound GdN. In a first step, we discuss and study the ground state phase diagrams explored in the different planes of physical parameter. The obtained stable configurations are illustrated by comparing the energies modeled by the Hamiltonian of Eq. (1). For non-null temperature values, we present and discuss the behavior the total magnetizations and susceptibilities as a function of: the temperature, the crystal field, the exchange coupling interactions and the external magnetic field. The hysteresis loops are illustrated and analyzed when varying the external magnetic field for specific values of the physical parameters: the temperature, the crystal field and the exchange coupling interactions. In order to complete this study, we present the obtained values of the critical exponents corresponding to compound GdN. Such results have been compared with those existing in the literature.

**KEYWORDS:** Rare earth GdN alloy; Monte Carlo; Hysteresis loops; Magnetic properties; Critical exponents.

*) Corresponding authors: lahou2002@gmail.com(L,B) ; samiraidrissi2013@gmail.com (S,I);

## I. Introduction

In the last decade, Gadolinium nitride and related rare-earth nitrides have been intensively studied, due to their interesting applications in the field of semiconductor spintronics [1, 2]. Also, these include GdN based spin-filter Josephson junctions [3, 4], field effect transistor structures [5]. In fact, such compound combine the properties of both semiconductors and ferro-magnets, so-called intrinsic ferromagnetic semiconductors, the several groups have already succeeded in obtaining proof of concept GdN based device structures. The use of GdN quantum dots enhanced the efficiency of GaN tunnel junctions [6] and the development of topological insulator GdN hetero-structure devices [7]. Moreover, the rare-earth nitrides combine both technologies: spin-filter and Josephson junctions because of containing structural properties compatible with the well-established group-III-nitride family. The lattice between AlN and either GdN or GaN of ~þ10%, and ~þ13% respectively, though relatively severe, is comparable to that in hetero-epitaxial systems [8]. Even the preliminary results have been proved [9–11]. Several theoretical and experiments calculations of GdN have been made during the past few years using ab-initio calculations to get an insight into their physical properties and to discuss the obtained results [12–23]. A challenge between both experiment and theory have been outlined regarding the rare-earth nitrides. Recent studies about GdN regarding whether it is an insulator, a metal, or a semiconductor, even if most RE nitrides have been revealed to be semi-metallics. Because of their partially filled 4f shells, an accurate description of the electronic structure of RE compound is a problem difficult to solve. Due to the open shell 4f states, GdN and other rare earth nitrides showed interesting electronic and magnetic properties. However, there exists divergences among several theoretical treatments concerning the nature of the calculated the electronic properties and magnetic behavior [23]. At low temperature, there are reports which illustrate degenerately doped semiconducting ground state based on the obtained resistivity data, see Ref. [24–26]. A conflict between the magnetic properties of the compound GdN with earlier observations have been described the metal-magnetic material [27], while other works revealed it to be a ferromagnetic behavior [23, 28]. Srivastva et al. [29] using the muffin-tin orbital method in the framework of the LDA approximation, to calculate the structural, the electronic and the magnetic properties of the GdN alloy under pressure. Sharma and Nolting affirmed that the electronic properties allowed and understand the magnetism and analyzed the possibility of carrier-mediated ferromagnetism in GdN alloy [30]. Duan *et al.* establish that the GdN displays ferromagnetic ground state with a half-metallicity gap of about 0.6 eV when using the first principles studies including the electronic and magnetic properties of GdN alloy [31].

Lambrecht et al. showed that the frozen phonon calculation for several rare earth nitrides counting GdN based on the LSDA + U approximation, and the full-potential linearized muffin-tin orbital technique. However, highlighted the necessity for the computation of the full phonon dispersion curve in the Brillouin zone and phonon density of states. They also reported that phonon frequencies appeared only at some high-symmetry points of the Brillouin zone [24]. Khazen et al, reported a single peak ferromagnetic resonance in both the in plane and out of plane GdN epitaxial films exist. While, its lattice constant is found to be in the interval [5.01–5.12] Å [32]. Among very few IFSs, the GdN alloy is mainly interesting because the Gd element has a half-filled 4f shell which determines the interatomic 4f↔5d exchange interaction of the individual ions. Thus 4f shell contributes with a large magnetic moment of $Gd^{3+}/\mu 4f$=7.0 $\mu_B$ [33–36]. The material GdN contains a strong magnetic moment, this is why we have motivated to study the magnetic properties of this compound using Monte Carlo Simulation Method, in addition we have already treated the magnetic properties with this method see the [37-43]. In this paper, we started with the ground state study. For this purpose we propose a Hamiltonian describing our system. For non-null temperature values, we performed Monte Carlo simulations under the Metropolis algorithm. This paper is organized as follows: in section II, we present the model and the Hamiltonian describing the compound GdN. In section III, we present and discuss our Monte Carlo simulations. Section IV is devoted to conclusions of the obtained results.

## II. The Hamiltonian model of the GdN alloy

The magnetism of the GdN compound comes from the gadolinium ($Gd^{3+}$) with large magnetic moment $\mu_{eff}^{theor}$=7.0 $\mu_B$ / $Gd^{+3}$ and a magnetic spin moment S=7/2. The geometry of the studied system is belongs to the space group Fm-3 m (No. 252). Such geometry is presented in Fig.1, using VESTA software [44].

**Fig.1**

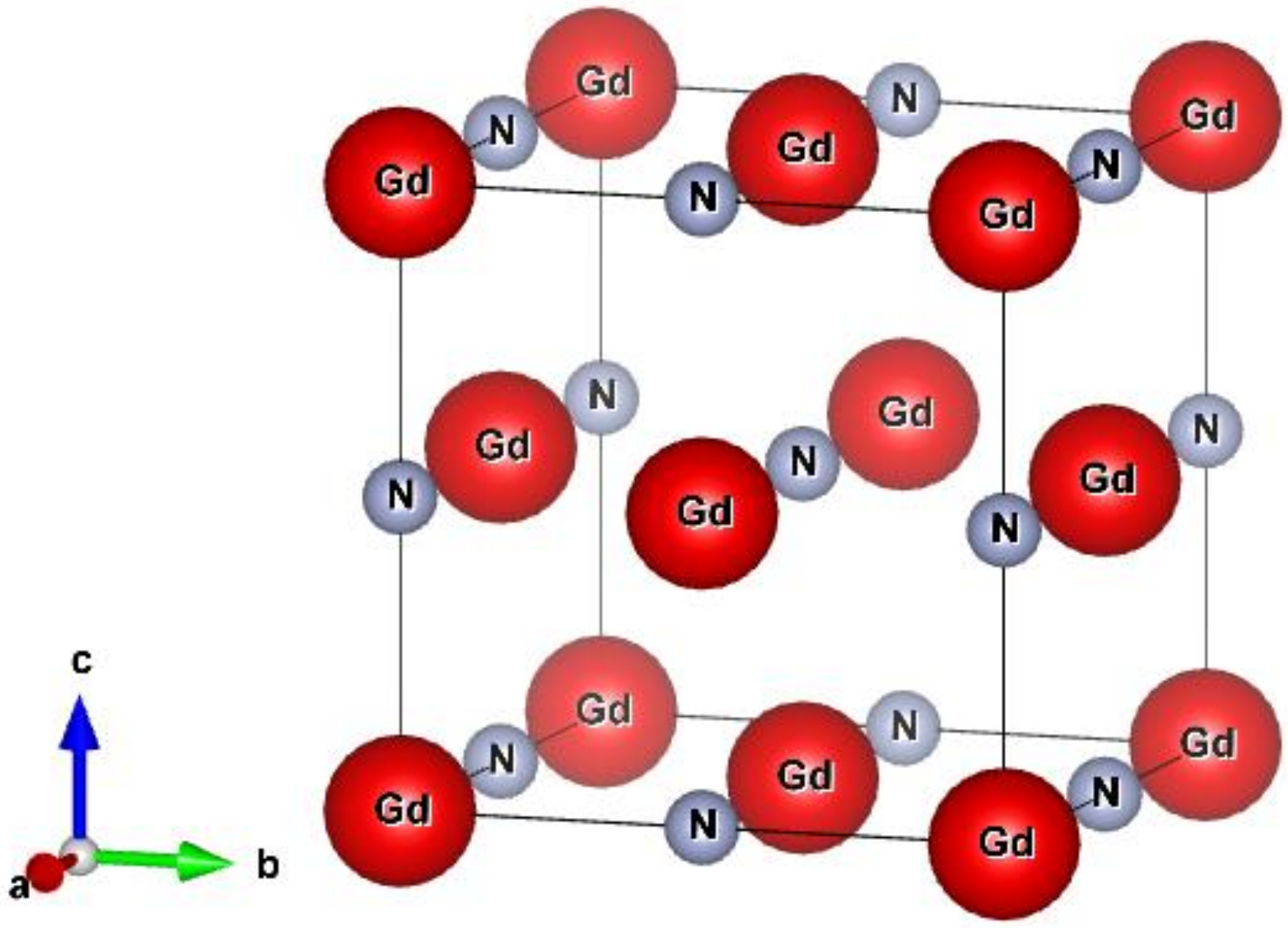

Fig.1: Unit cell of the GdN structure using Vesta software [44].

The Hamiltonian modeling this system is as follows:

$$\mathcal{H} = -J_1 \sum_{<i,j>} S_i S_j - J_2 \sum_{<i,k>} S_i S_k - H \sum_i S_i - D \sum_i {S_i}^2 \quad (1)$$

Where the spin variables are: $S_i=\pm 7/2$, $\pm 5/2$, $\pm 3/2$ and $\pm 1/2$.

$J_1$, $J_2$ are for the first and the second exchange coupling interactions between the Gd atoms, respectively. H is the external magnetic field and D is the crystal field of Gd ions.

## III. Monte Carlo simulation of the GdN alloy

In this section, we study and illustrate the magnetic properties of the GdN compound. For this purpose, we perform Monte Carlo simulations via the Metropolis algorithm. This method is founded on the Markov process in the bases of the Hamiltonian of Eq. 1. In a first step, different initial configurations are used to reach the equilibrium of the studied system GdN. For each spin arrangement, a number of $10^5$ Monte Carlo steps are achieved; eliminating the first $10^4$ generated configurations. During each Monte Carlo step, the total sites of the system are visited and a single-spin flip attempt is completed. The Metropolis algorithm accepts or rejects the changes made in the spin configurations. The free boundary conditions are realized for the finite lattice size. For the calculations of the ground state phase diagrams, we have fixed the system size: N=5x5x5. The scheming of the physical parameters such as the magnetization, the magnetic susceptibility, the specific heat and the energy of the studied system are done. The physical quantities are estimated via the following expressions:

The total energy per site of the GdN system is:

$$E_{tot} = \frac{1}{N} < \boldsymbol{\mathcal{H}} > \quad (2)$$

The total magnetization of the system:

$$M_{tot} = \frac{1}{N} < \sum_i S_i > \quad (3)$$

The total susceptibility is:

$$\chi_{tot} = \frac{\beta}{N}\left(\langle {M_{tot}}^2 \rangle - \langle M_{tot} \rangle^2\right) \quad (4)$$

The total specific heat is:

$$c_v = \frac{\beta^2}{N}\left(\langle {E_{tot}}^2 \rangle - \langle E_{tot} \rangle^2\right) \quad (5)$$

With, $\beta = \frac{1}{K_B T}$, where the Boltzmann constant $K_B$ is fixed to its unit value, T is the absolute temperature.

The critical exponents: α, β and γ of the studied system are determined by [45–47] :

$$m \sim |T_C - T|^{\alpha} \quad (6)$$

$$\chi \sim |T_C - T|^{\beta} \quad (7)$$

$$C_V \sim |T_C - T|^{\gamma} \quad (8)$$

These critical exponents are calculated near the critical temperature.

***Table 1. The Critical temperature $T_c$ of the compound GdN.***

| | Experimental studies | Present study (Monte Carlo) |
|---|---|---|
| $T_c$ | 30 K [48]<br>50 K [49] | 33.5 K |

## A. Ground state study

Using the Hamiltonian of the Eq. (1), we calculate the different energy values corresponding to the different possible configurations: 2S+1=8 with S=7/2. The most stable configurations correspond to the minimum of the calculated energy.

In order to illustrate the ground state phase diagrams in different planes, we present in Figs. (2a, 2b, 2c and 2d) the corresponding stable phases for different physical parameters. In fact, Fig. 2 (a) shows the stable phases in the plane (H, D) for $J_1= J_2=1$. From this figure, it is found that the all eight possible: ±7/2, ±5/2, ±3/2 and ±1/2 are found to be stable in this plane. A perfect symmetry is found according to the axis H=0. For D>-10, the phases ± 3/2 are missing in this phase diagram. We illustrated Fig.2b in the plane ($J_1$, D) for H=0 and J2=1, the only two stable phases, namely: -1/2 and -7/2. To show the effect of varying the exchange coupling interaction $J_2$ on the stability of different configurations, we present in Fig.2c plotted in the plane (J2, D) the stable configurations for H=1 and $J_1$=1. From this figure it is seen that only three stable configurations in this figure, namely: -1/2, -3/2 and -7/2. The phase -3/2 is stable only in the region below the equation $J_2$ = -4.14 x D - 30.39, whereas the other phases are stable for any values of the parameters $J_2$ and D. Similarly to Fig.2c, we plot in Fig.2d, plotted in the plane ($J_1$, $J_2$) for D=H=1, the obtained stable phases. The same stable phases obtained in the Fig.2c are still present in Fig.2d, the only difference is that the phase -1/2 is stable only in the region below the equation $J_2$ = -10.41 x $J_1$ – 24.15.

Fig.2a

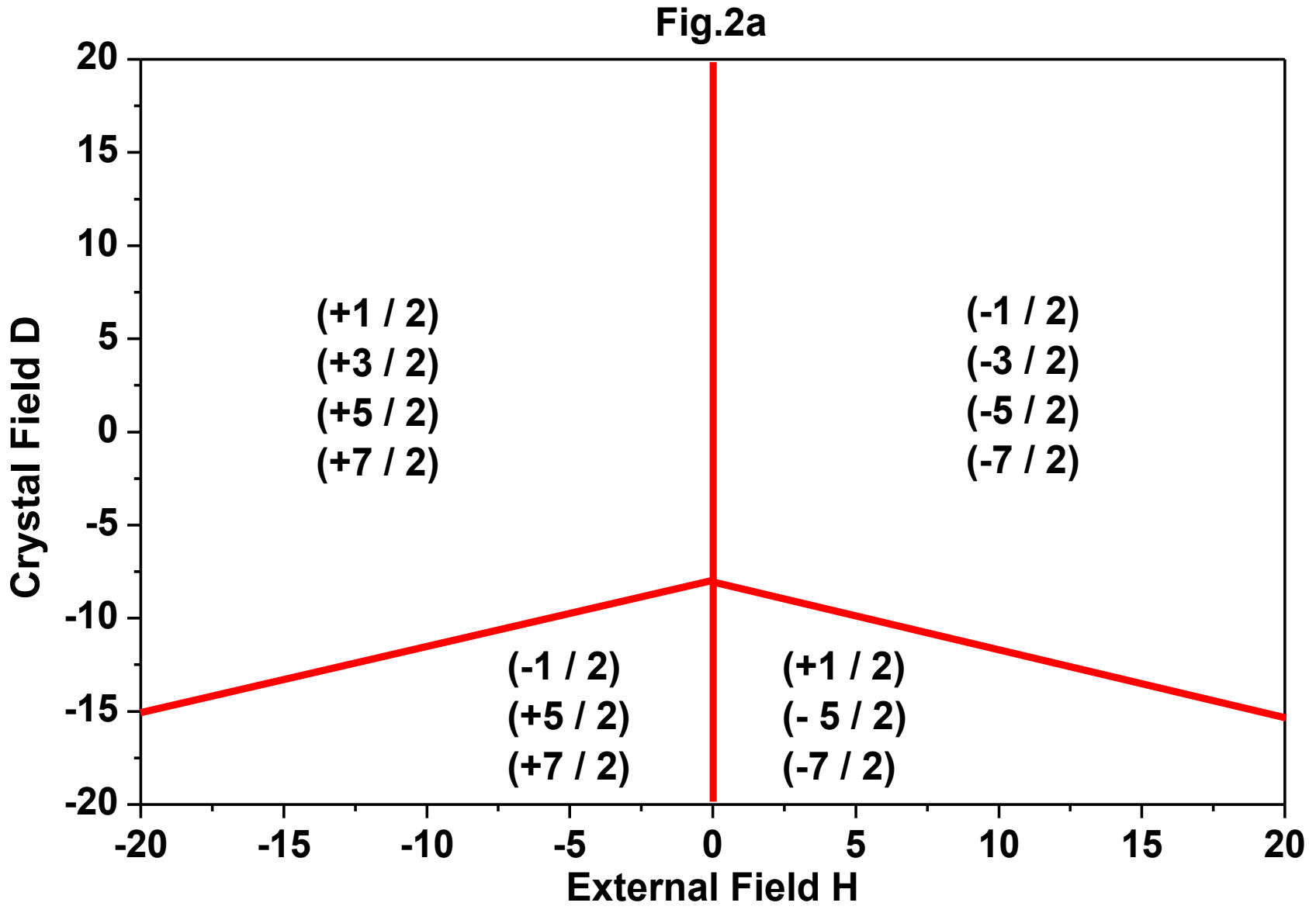

Fig.2b

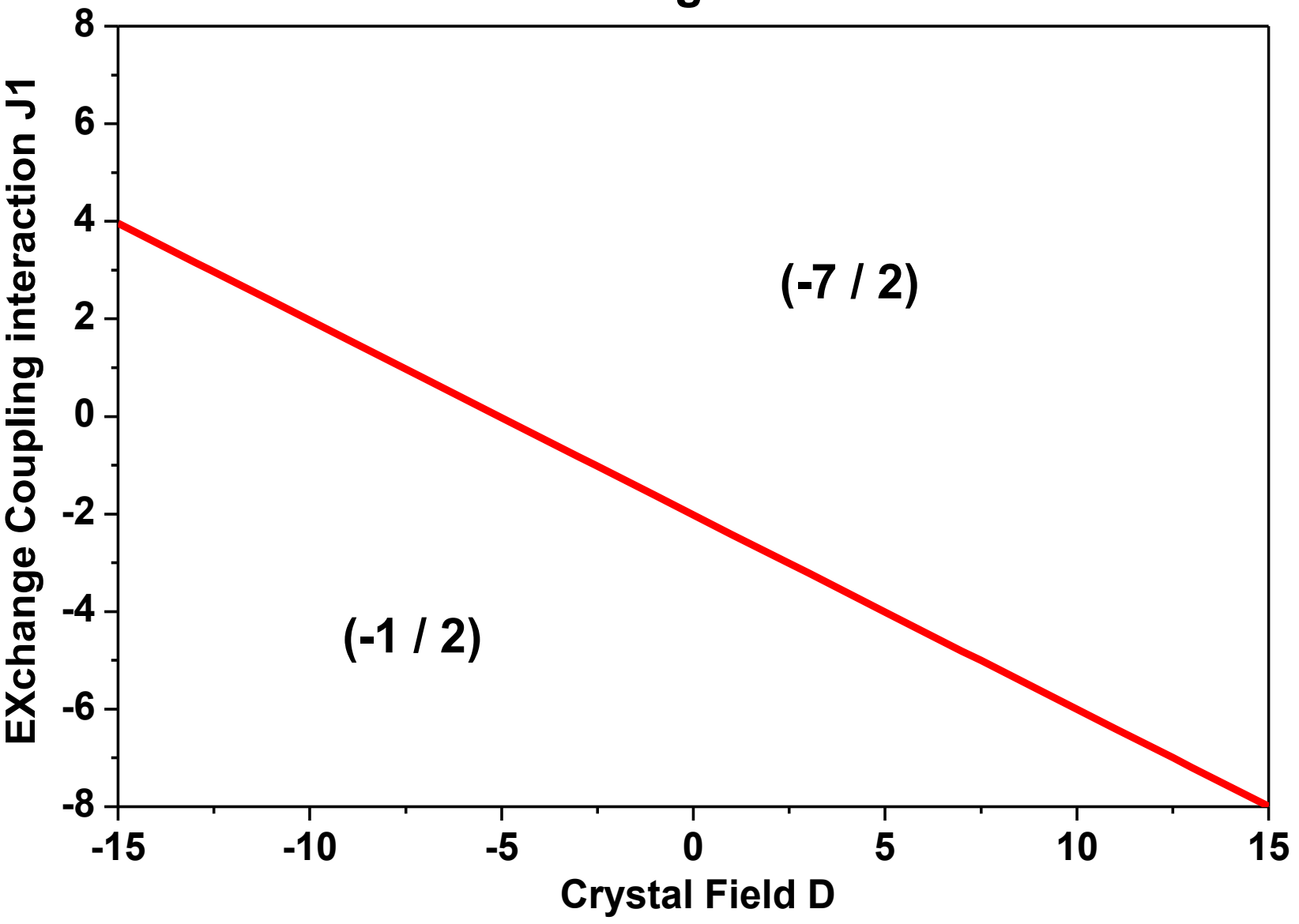

Fig.2c

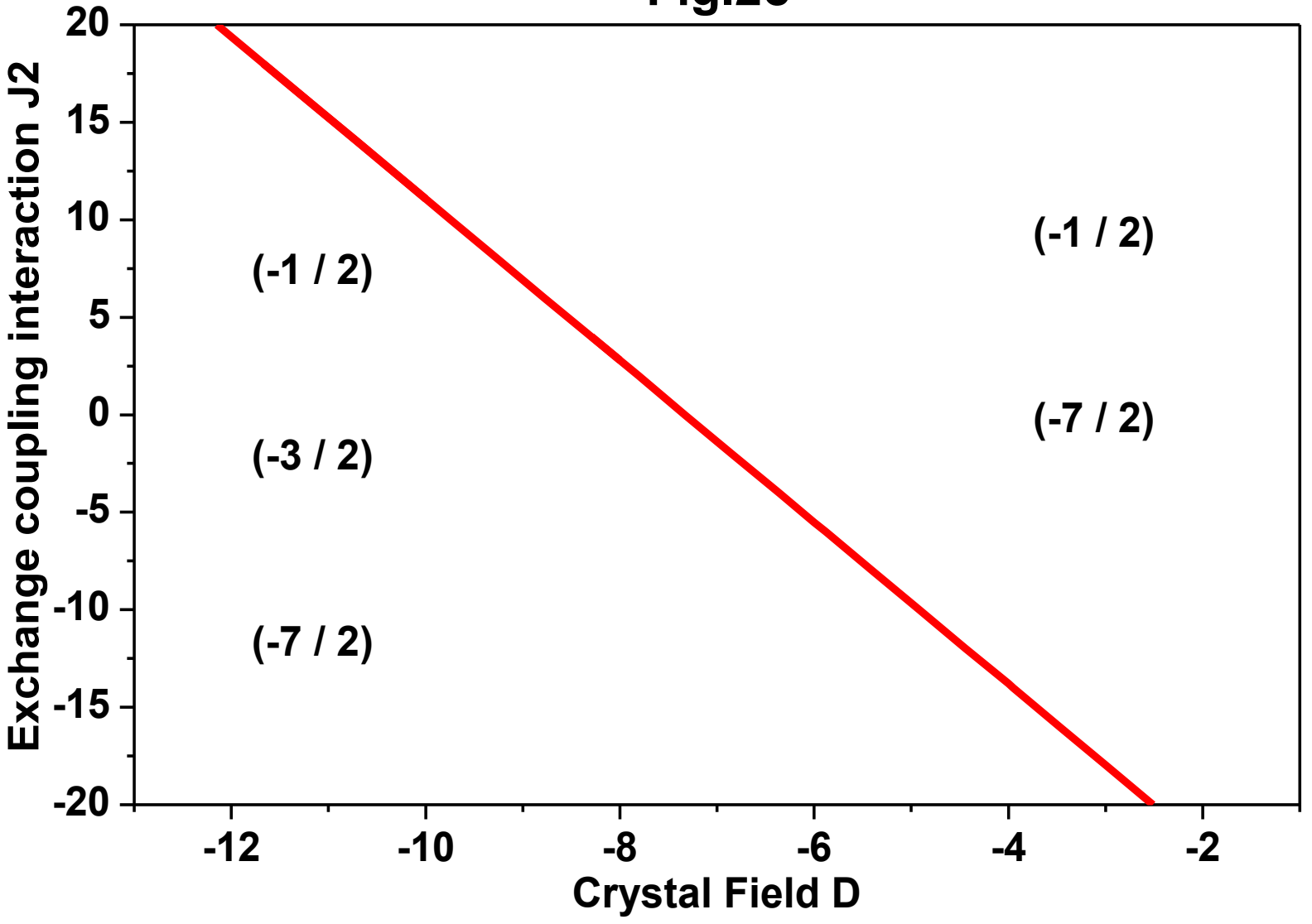

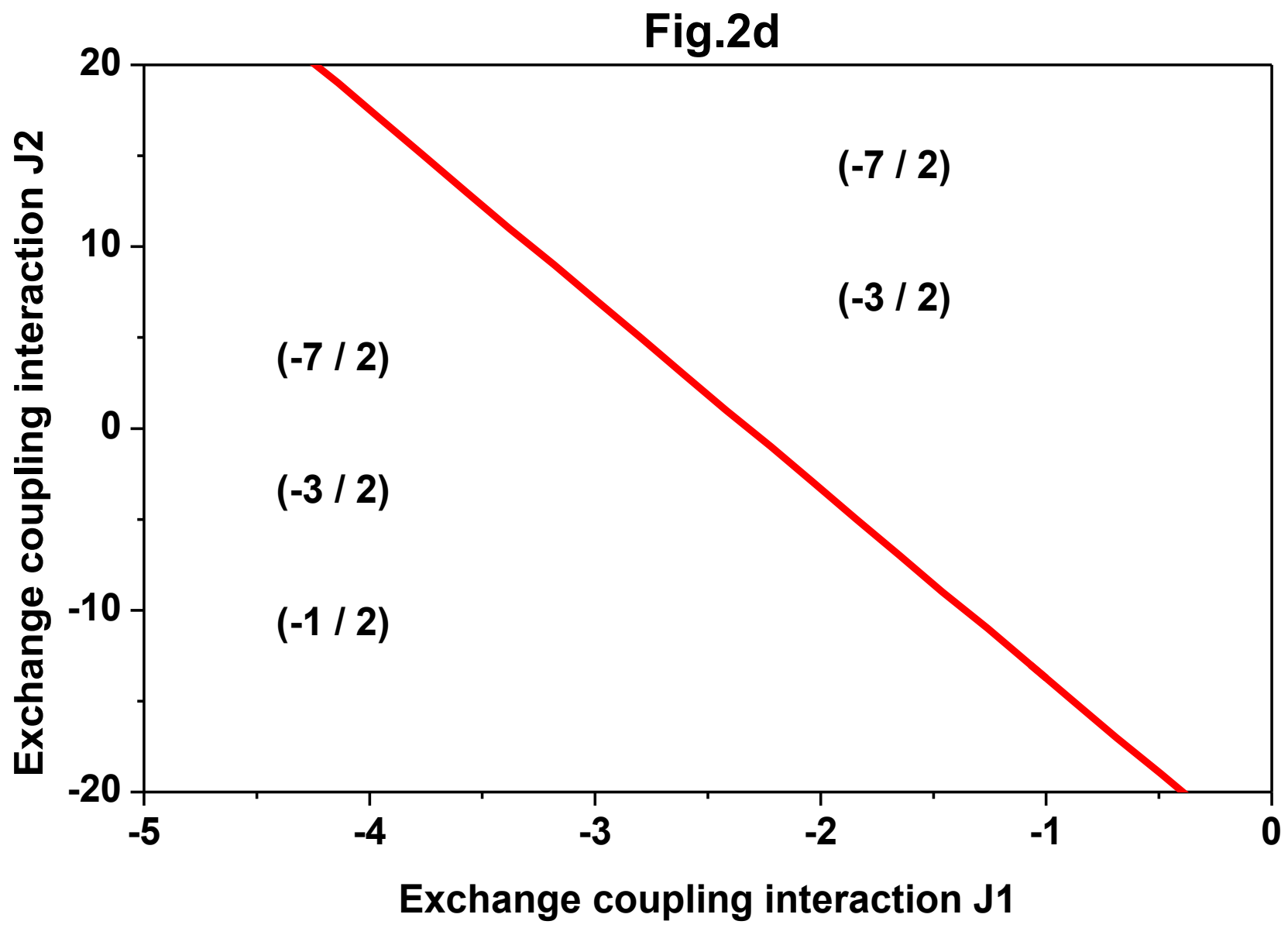

*Fig.2: Ground state phase diagrams: (a) in the plane (H, D) for $J_1 = J_2 = 1$, (b) in the plane (D, $J_1$) for H=0 and $J_2=1$; (c) in the plane (D, $J_2$) for H=1 and $J_1=1$, (d) in the plane (J1, $J_2$) for D=H=1.*

## B. Monte Carlo results of the GdN alloy

To perform Monte Carlo simulations under the Metropolis algorithm, we use the Hamiltonian describing this system, see Eq. (1). We begin by from different initial configurations to influence the equilibrium of the considered system. The Metropolis algorithm allows to accept or reject the different spin configurations. The free boundary conditions are imposed for the system.

To inspect the thermal behavior of the total magnetization for different system sizes, we plot in Fig.3a, the obtained results for H=1, $J_1=J_2=1$ and D=1. From this figure it is seen that for very low temperature values the behavior of the magnetizations reaches the value 7/2 of the ground phase diagrams, this is in good agreement with in Fig.2a. Also, when increasing the system size the temperature decreases, this is due to the size effect when varying this size from 2x2x2 to 8x8x8. In relationship with Fig.3a, we plot in Fig.3b the specific heat, plotted for the same physical parameters and the same system sizes. From this figure, we found that the peak of the specific heat increases, when the system size decreases. Likewise, the corresponding transition temperature increases when increasing this system size. To examine the behavior of the susceptibility of the studied system, we plot in Fig.3c the obtained results for H=1, $J_1=J_2=1$ and D=+1. This figure is plotted a fixed system size 5x5x5, showing that this parameter presents a peak at the temperature value Tc ≈ 33.5 K.

To complete this study, we provide in Fig.3d the behavior of the total energy for the same parameter values as a Fig.3c. The total energy increases with increasing temperature values then reaches its saturation for large values of this parameter.

The behavior of the magnetizations as a function of the crystal field is presented in Fig.4a for H=1, $J_1$=$J_2$=1 and selected values of temperature T=5 K, T=35 K, T=50 K and T=100 K. For all temperature values, the effect of increasing crystal field values is to pass into the system from the paramagnetic phase to the ordered phase having the saturation value $M_{tot}$= 7/2. This transition is found for the crystal field value D=-5. The effect of varying the exchange coupling interaction on the behavior of the total magnetization is plotted in Fig.4b for H=0, T=35 K, and selected values of the exchange coupling interactions. From this figure it is found that the increasing crystal field effect is to pass the system from the disorder phase to the ordered phase except for $J_1$ with negative values and/or $J_2$ negative. As in Fig.4a, the transition is reached for the crystal field value D=-5.

When changing the exchange coupling interaction $J_1$, the conduct of the total magnetization is obtained in Fig.4c for $J_2$=1, T=35 K, H=1 and selected values of the crystal field D=+3, D=0 and D= -3. In the region $J_1$<-4, the system undergoes the paramagnetic phase ($M_{tot}$=0). For $J_1$>-4, the system reaches its saturations respecting the sign of the crystal field D. In fact, for D>0 the total magnetization reaches the value Mtot=3.5; while for D≤0 the total magnetization undergoes the value Mtot=-3.5.

The effect of growing the exchange coupling interaction $J_2$ on the performance of the total magnetization is illustrated in Fig.4d for the fixed values $J_1$=1, T=35K, H=1 and selected values of crystal field D=+3, D=0 and D= -3. The total magnetization reaches the saturation value $M_{tot}$=+3.5 for positive values of $J_2$. While the effect of increasing the crystal field is not felt.

The hysteresis loops of the studied system are plotted in Figs. (5a and 5b), when fixing the temperature value T=35 K. Fig.5a is plotted for $J_1$=$J_2$=1 and selected values of the crystal field D=-6, D=-2 and D=-1. From this figure it is seen that when increasing the crystal field, the surface of the loop as well as the corresponding coercive field increase. The effect of varying the exchange coupling interaction values on the comportment of the hysteresis loops is outlined in Fig.5b for D=0 and different values of the exchange coupling $J_1$ and $J_2$. From this figure it is found that the surface of the hysteresis loop is maximum for positive values of $J_1$ and $J_2$. While, for $J_1$ with a negative values a double loops is appearing in the hysteresis cycle. The antiferromagnetic behavior of the system is proved when imposing negative values of the exchange coupling interaction.

To study the critical behavior of the system GdN, we present in Figs.6 (a, b and c) the behavior of the critical exponents. In fact, Fig.6a represents log(m) as a function of log(|T-Tc|) with a slope equal to the critical exponent α=0.25. This value is in good agreement with that one found in the literature α=0.20, see Ref. [46].

In order to examine the critical exponent β behavior, we plot in Fig.6b: log (χ) as a function of log L. Our calculations lead to the value β=0.27, in good agreement with that one of the Ref. [46], which is β=0.30. To complete this study, we have investigated the critical exponent γ, when plotting log (Cv) as a function of log L. From this figure, we obtained the value γ=1.30 which is compared to that one given in Ref. [46] (γ=1.20).

Fig.3a

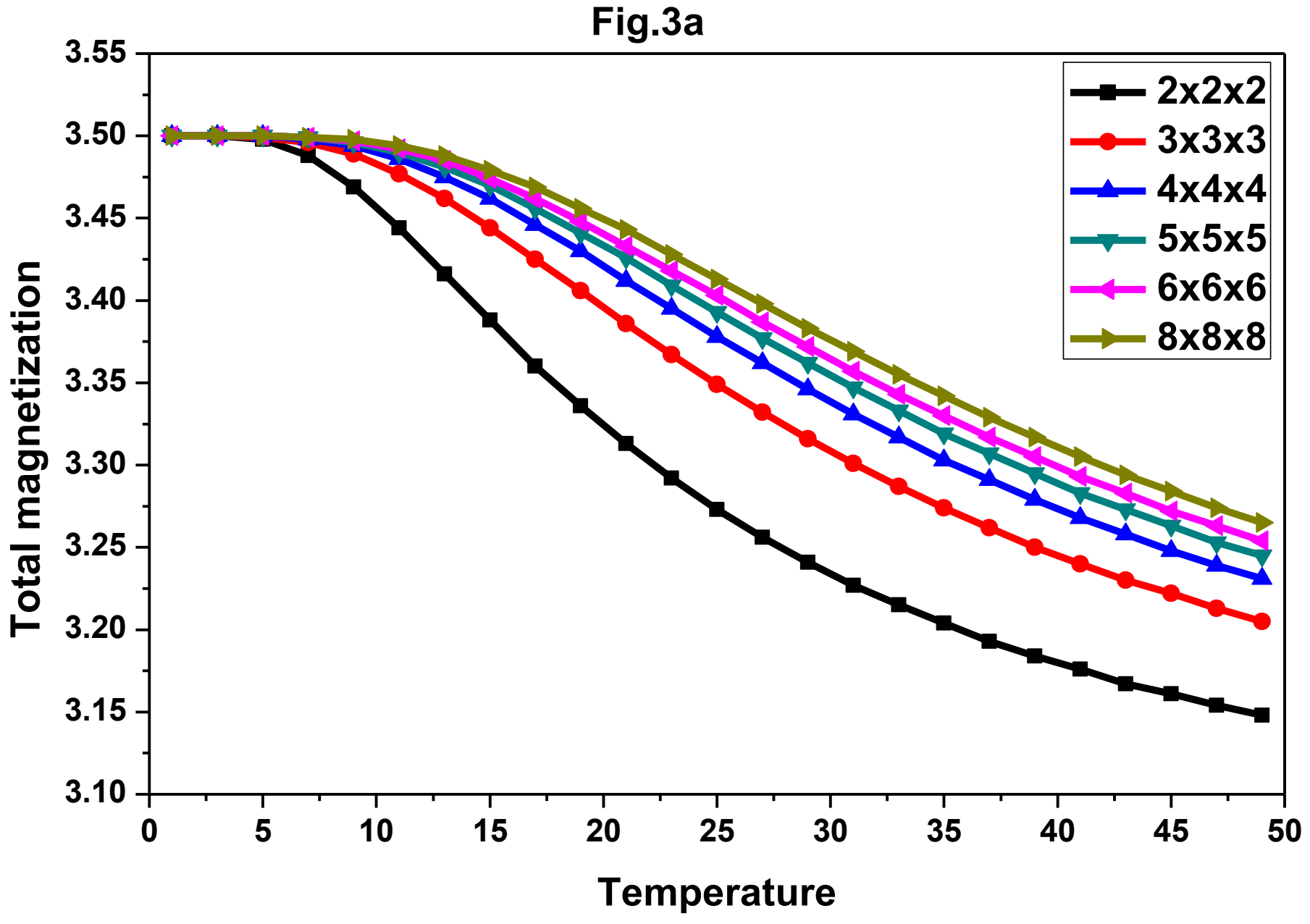

Fig.3b

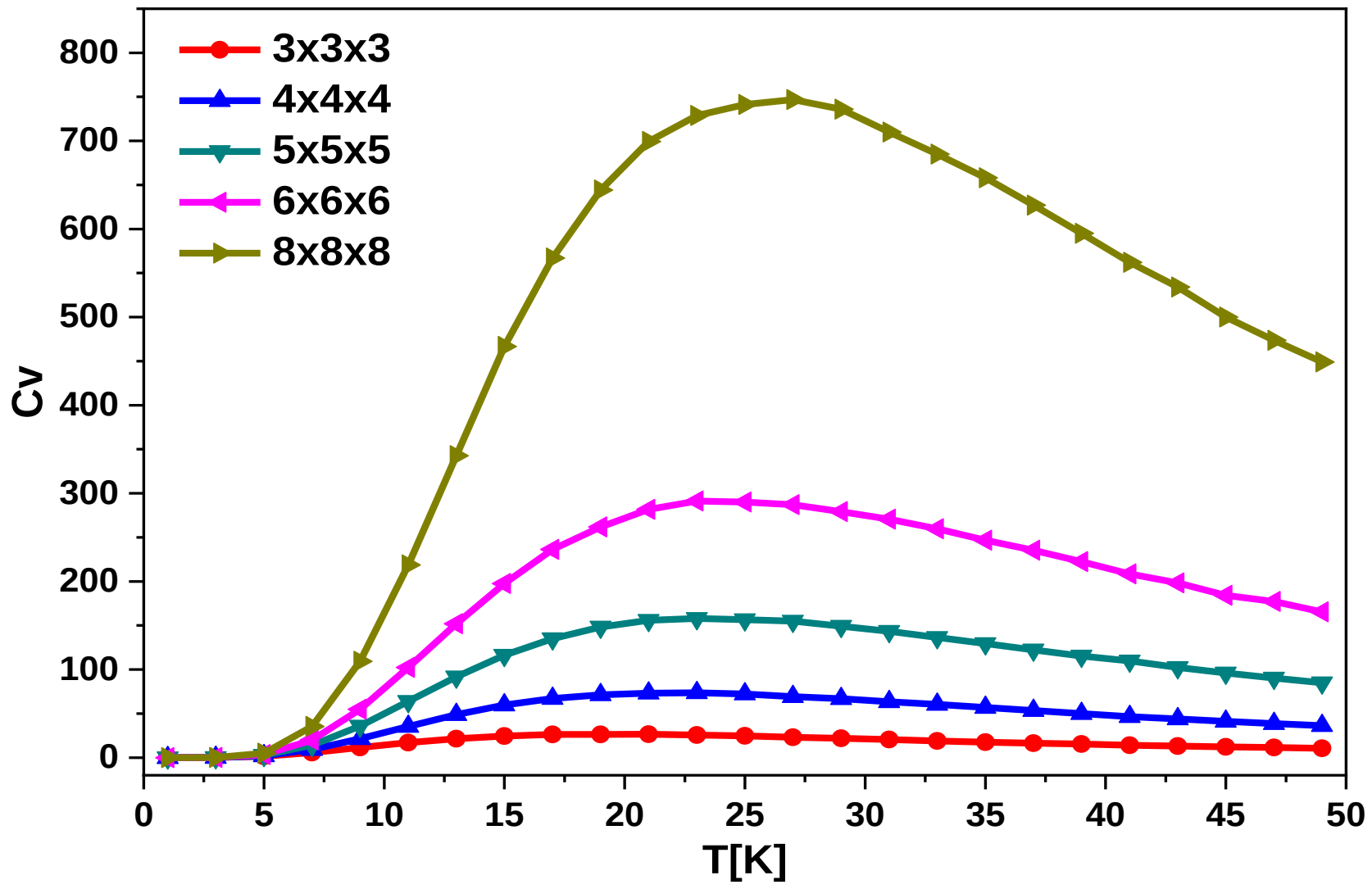
3x3x3
4x4x4
5x5x5
6x6x6
8x8x8
Cv
0
100
200
300
400
500
600
700
800
0 5 10 15 20 25 30 35 40 45 50
T[K]

Fig.3c

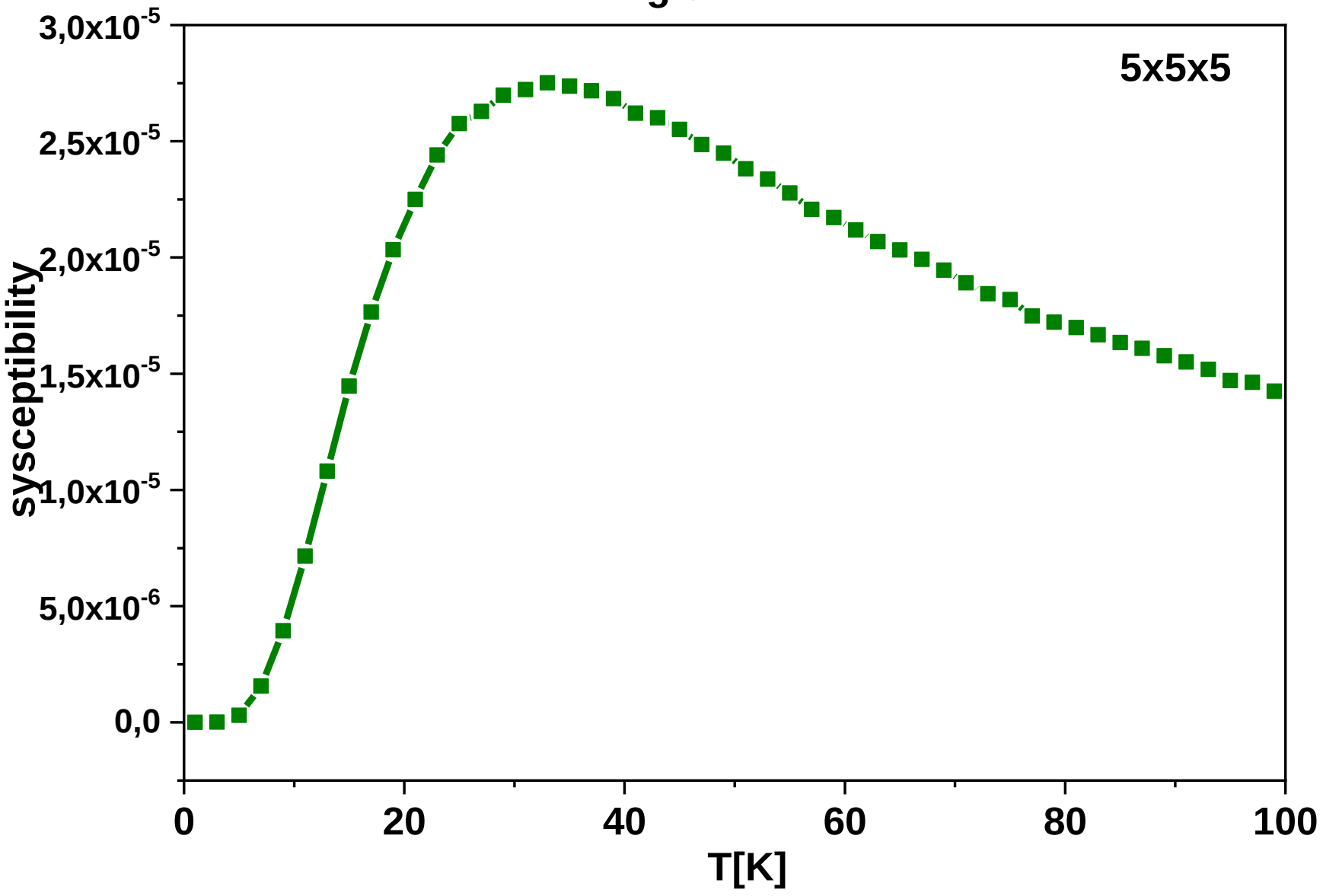

Fig.3d

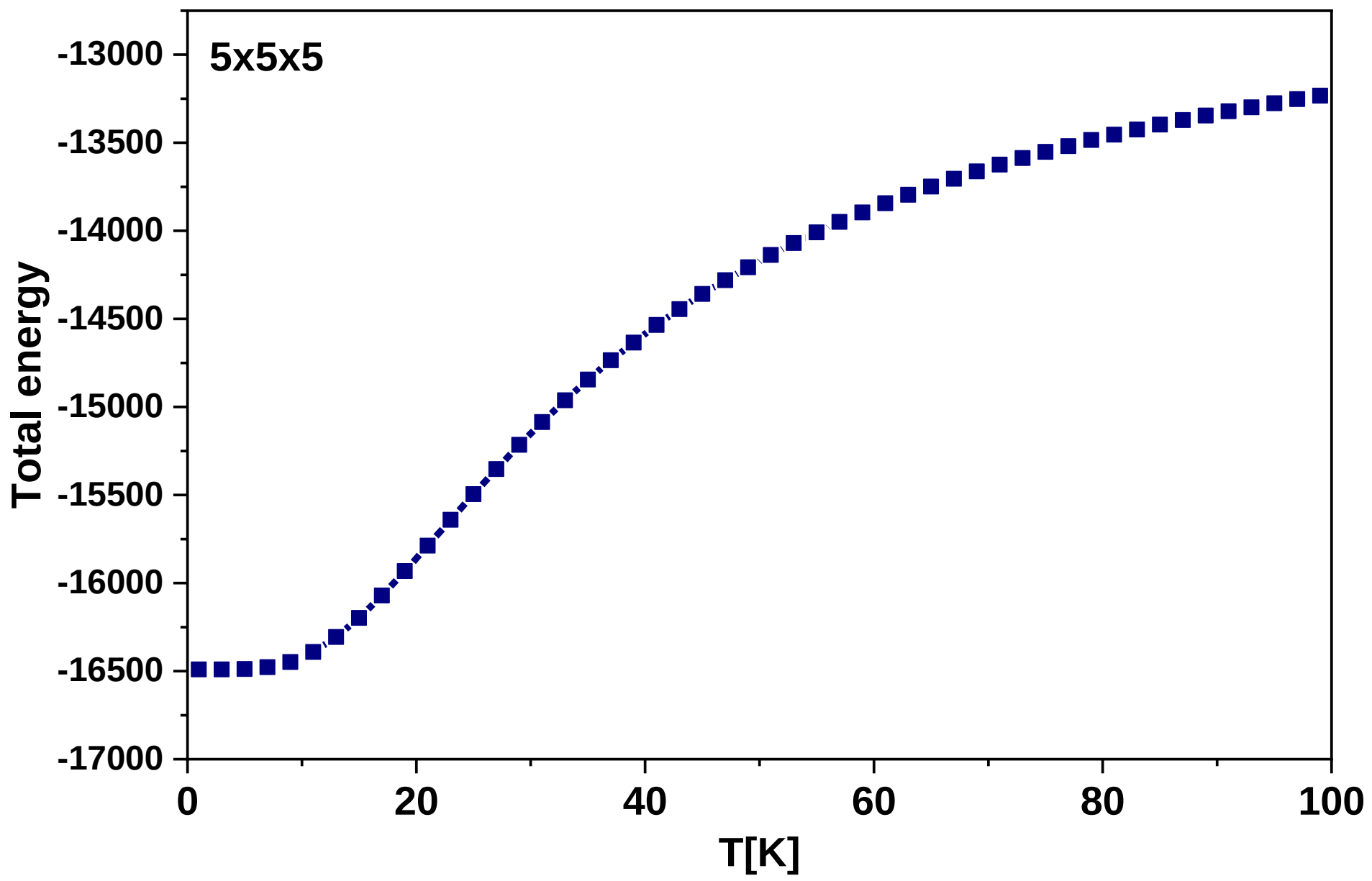

***Fig.3:** Thermal behavior of the physical parameters of the GdN alloy for H=1, $J_1=J_2=1$ and D=+1. In (a) we present the total magnetizations. In (b) we give the specific heat. In (c) illustrate the total susceptibility for the size* **5x5x5***. In (d) we present the total energy for the size* **5x5x5.**

Fig.4a

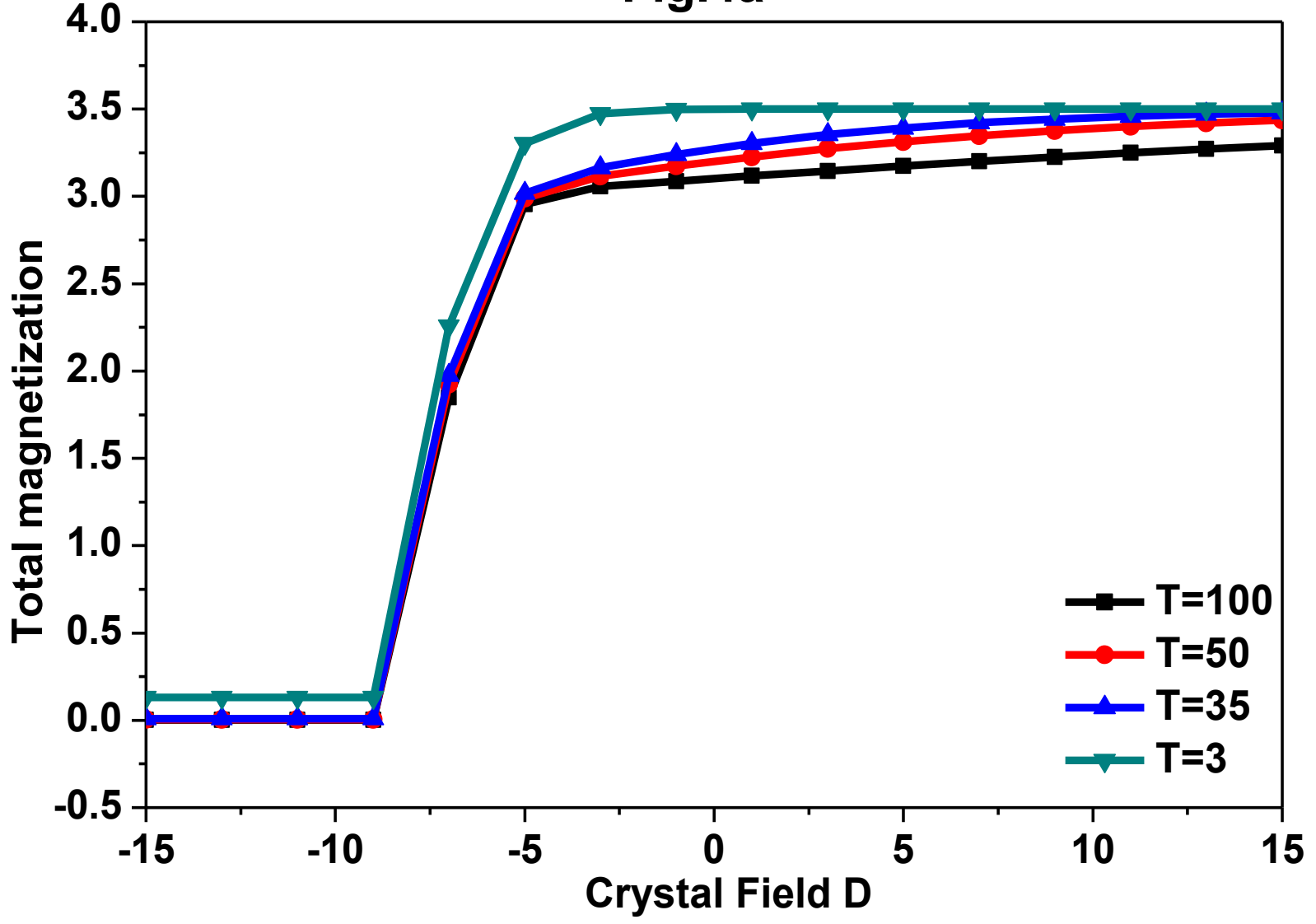

Fig.4b

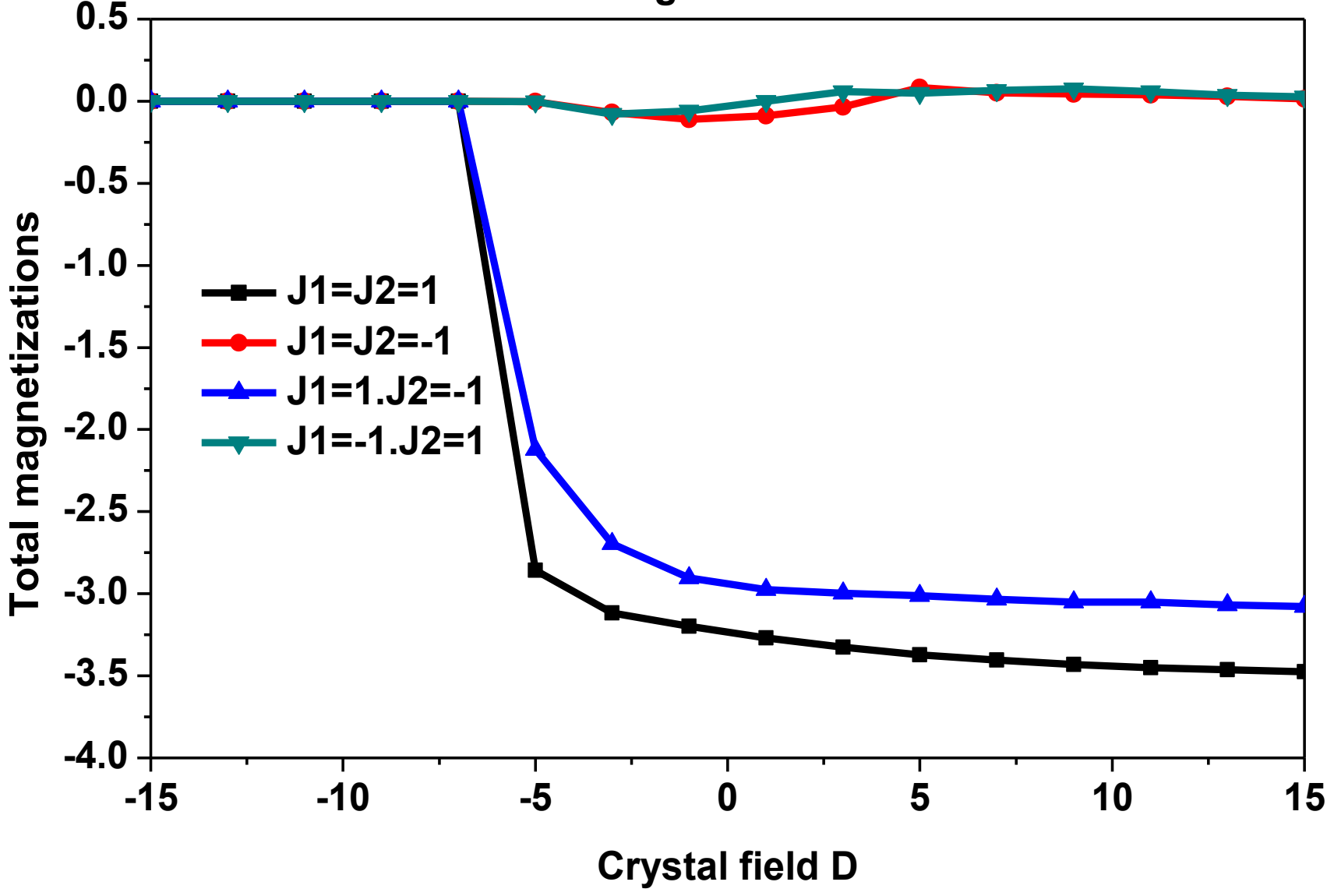

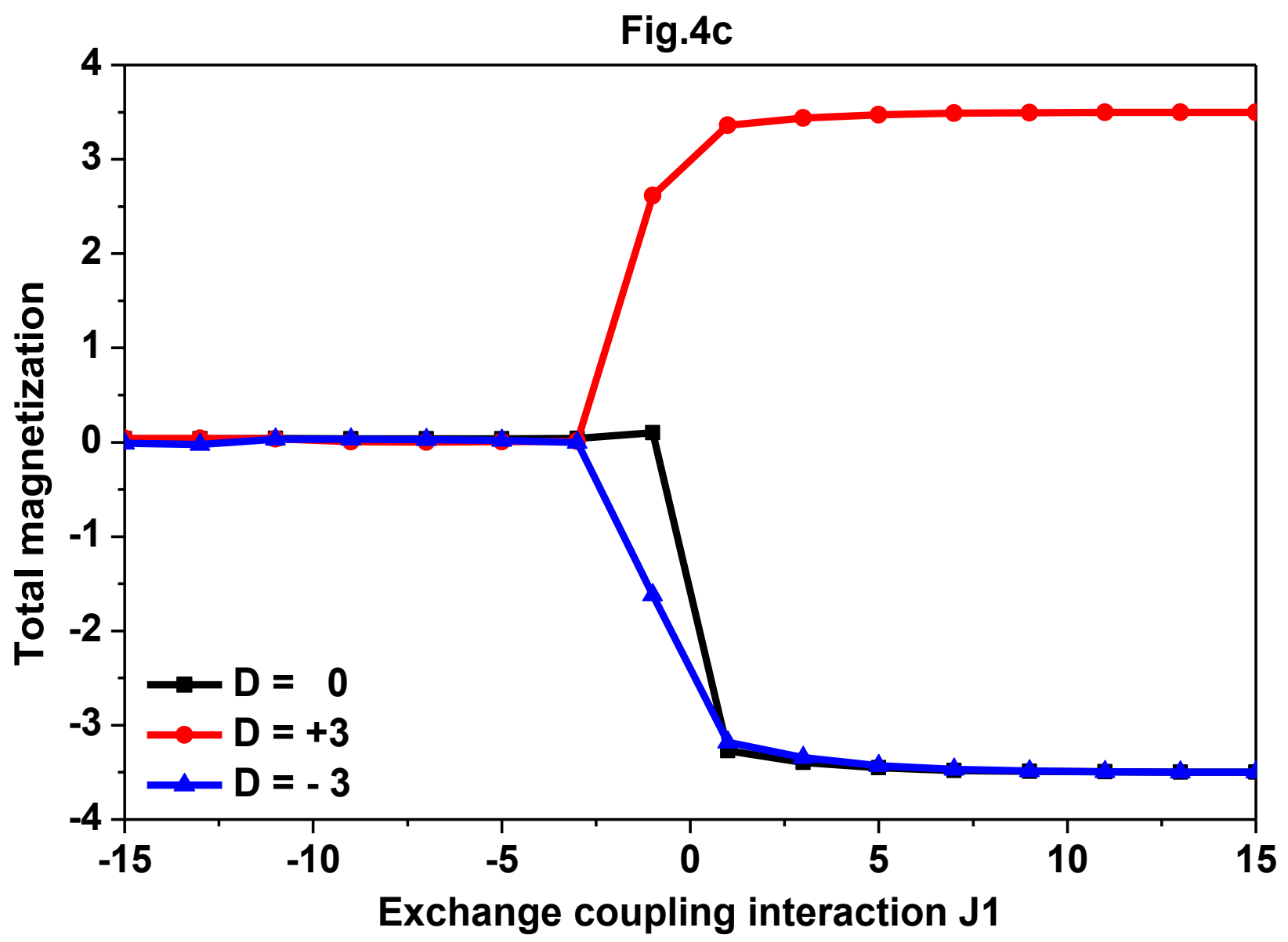

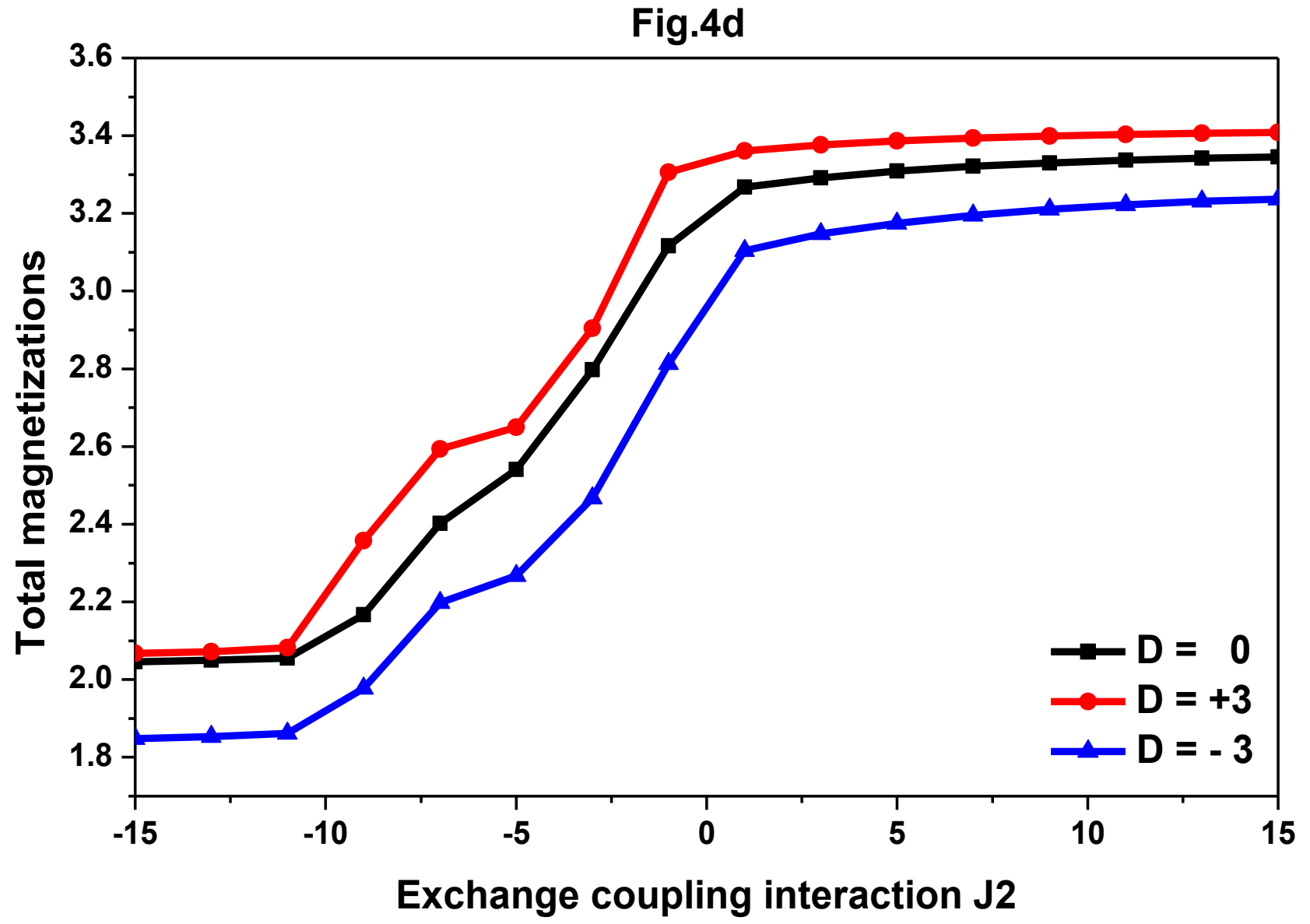

***Fig.4: Magnetizations of the studied system as a function of: In (a) the crystal field for H=1, $J_1$=$J_2$=1 and selected values of the temperature T=5 K, T=35 K, T=50 K and T=100 K. In (b) the crystal field for H=0, T=35 K, and selected values of the exchange coupling interactions. In (c) the exchange coupling $J_1$ for $J_2$=1, T=35K, H=1 and selected values of the crystal field D=+3, D=0 and D= -3. In (d) the exchange coupling $J_2$ for $J_1$=1, T=35K, H=1 and selected values of the crystal field D=+3, D=0 and D= -3.***

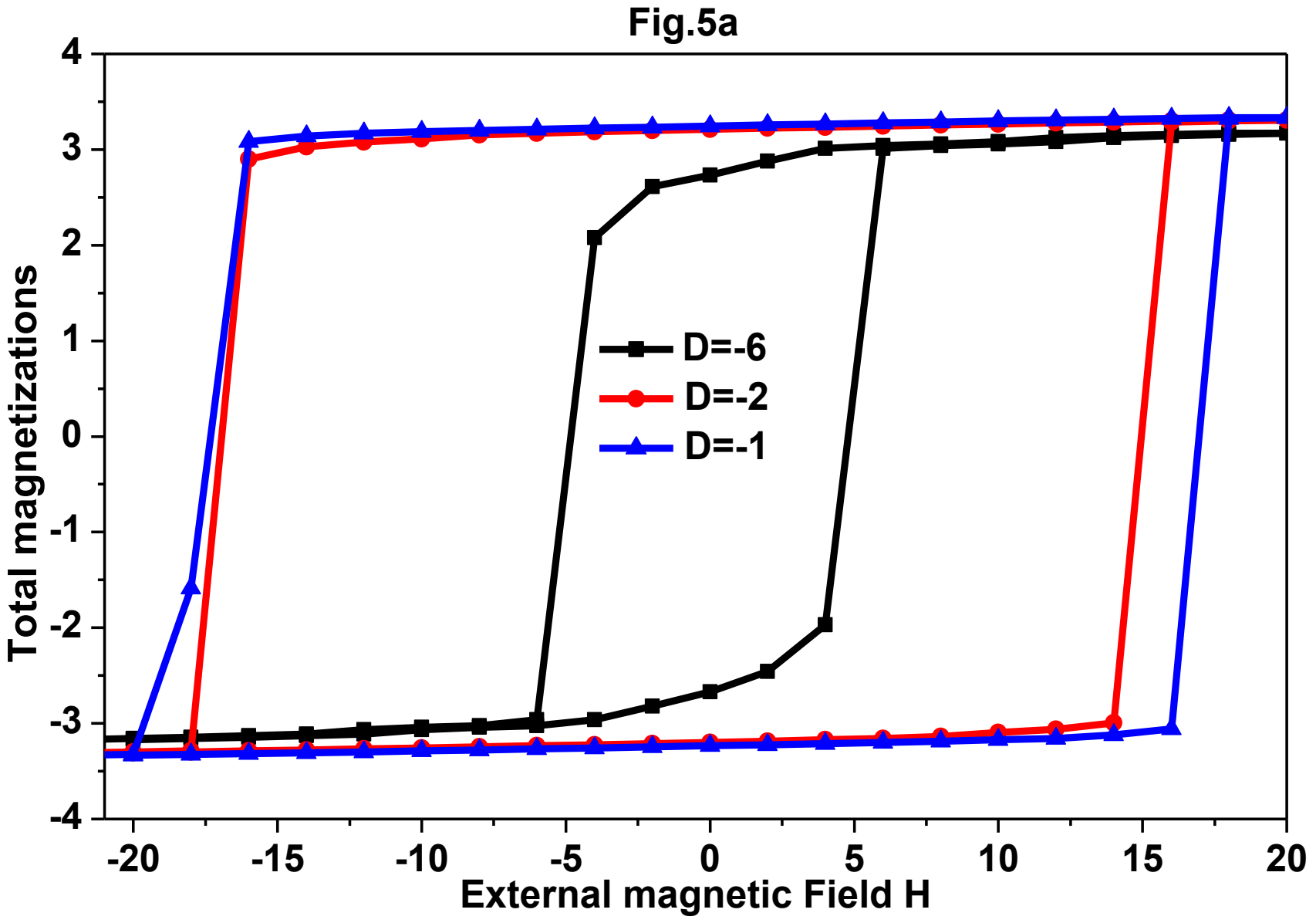

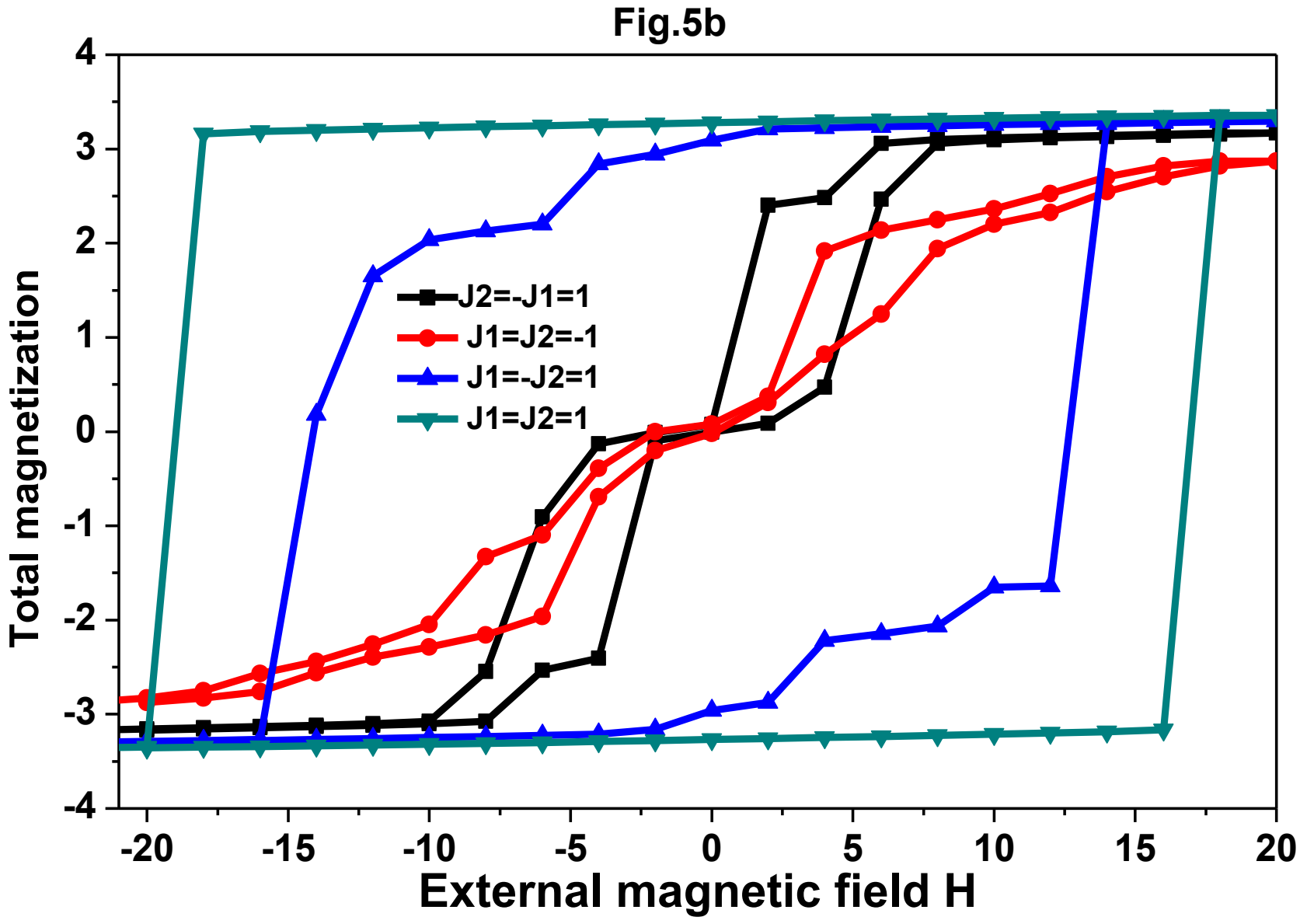

***Fig.5: Hysteresis loops of the compound GdN and specific value temperature T=35 K. In (a) for $J_1$=$J_2$=1 and selected values of the crystal field D=-6, D=-2 and D=-1. In (b) for D=0 and selected values of the exchange couplings $J_1$ and $J_2$.***

Fig.6a

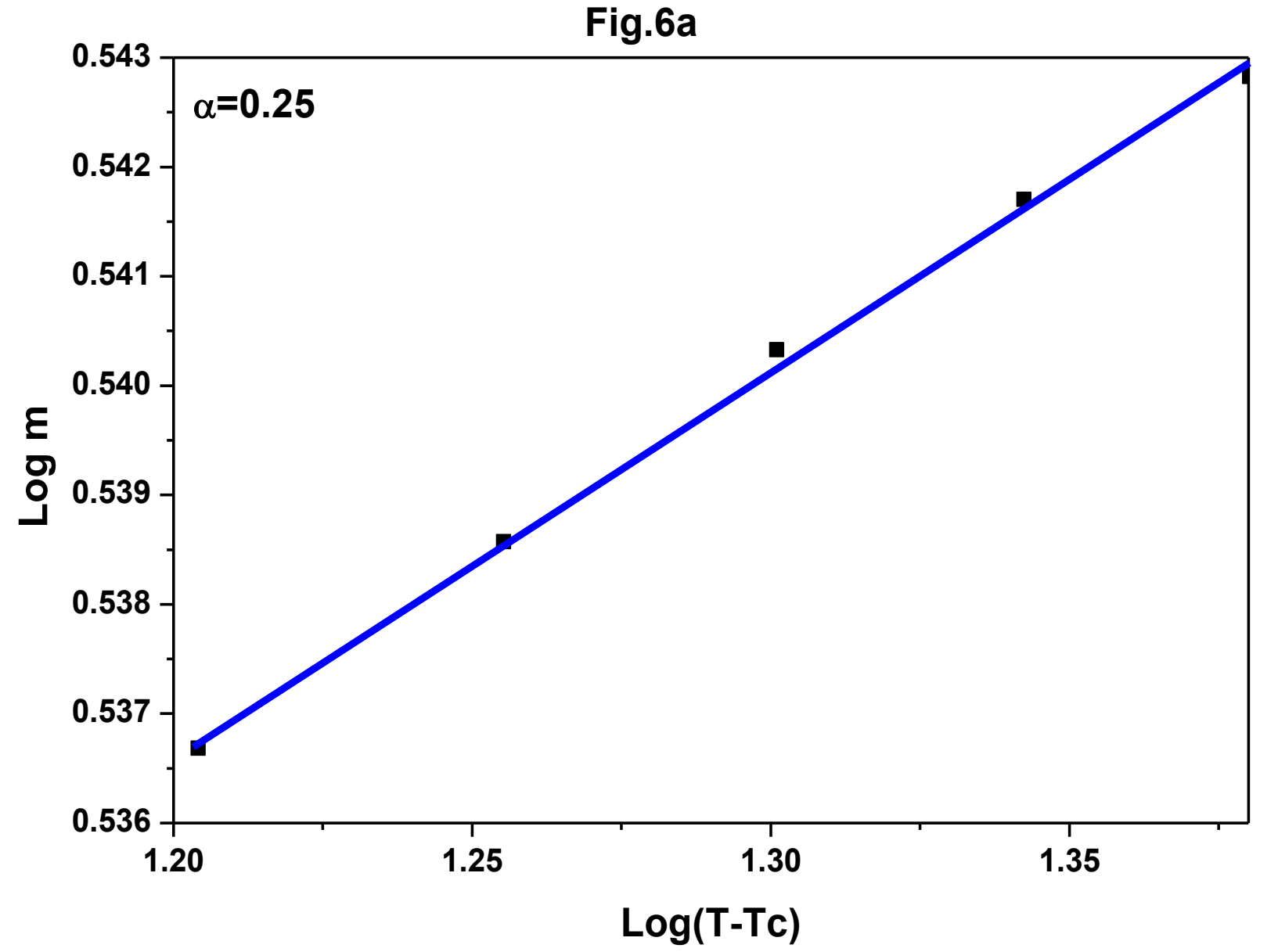

Fig.6b

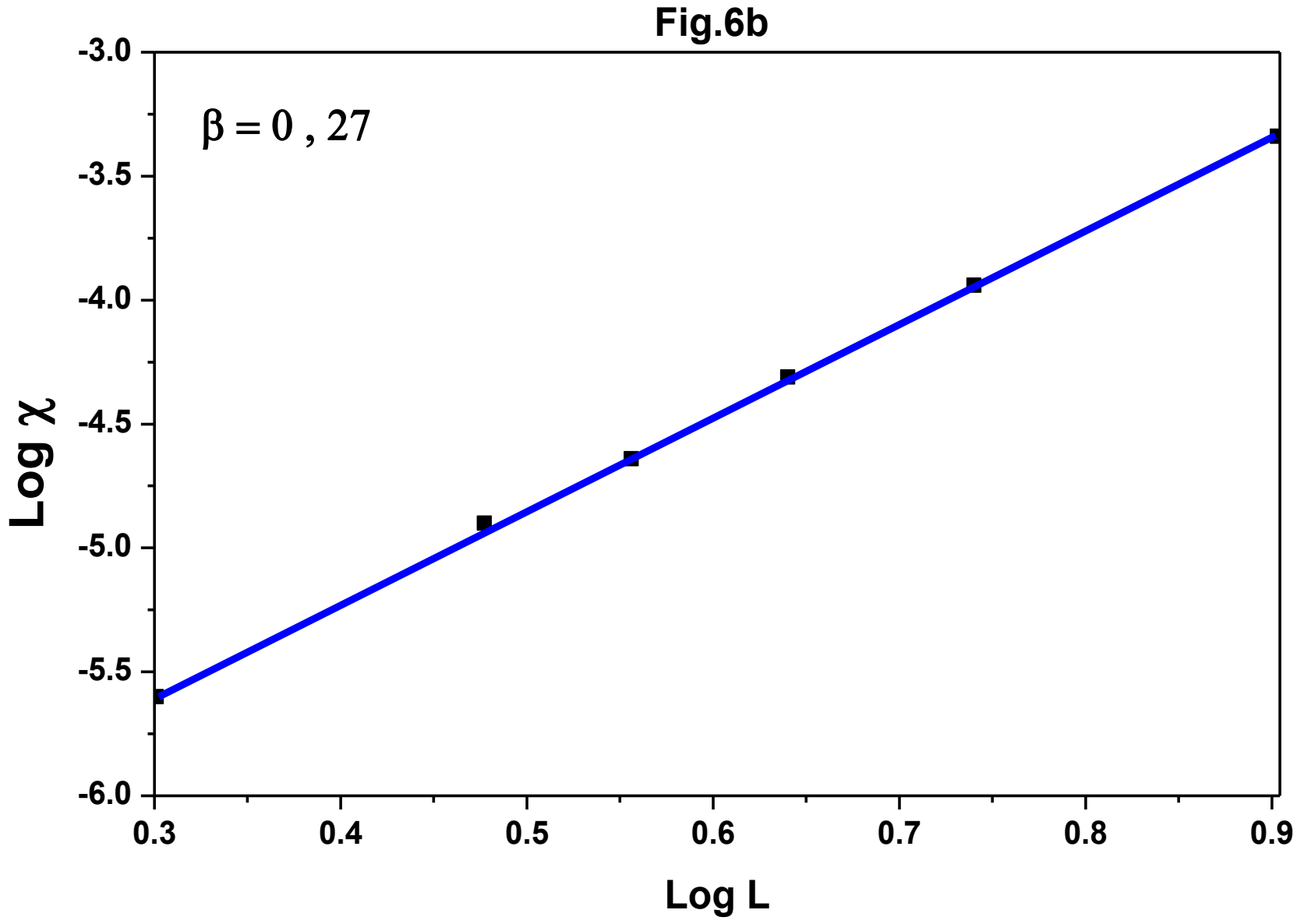

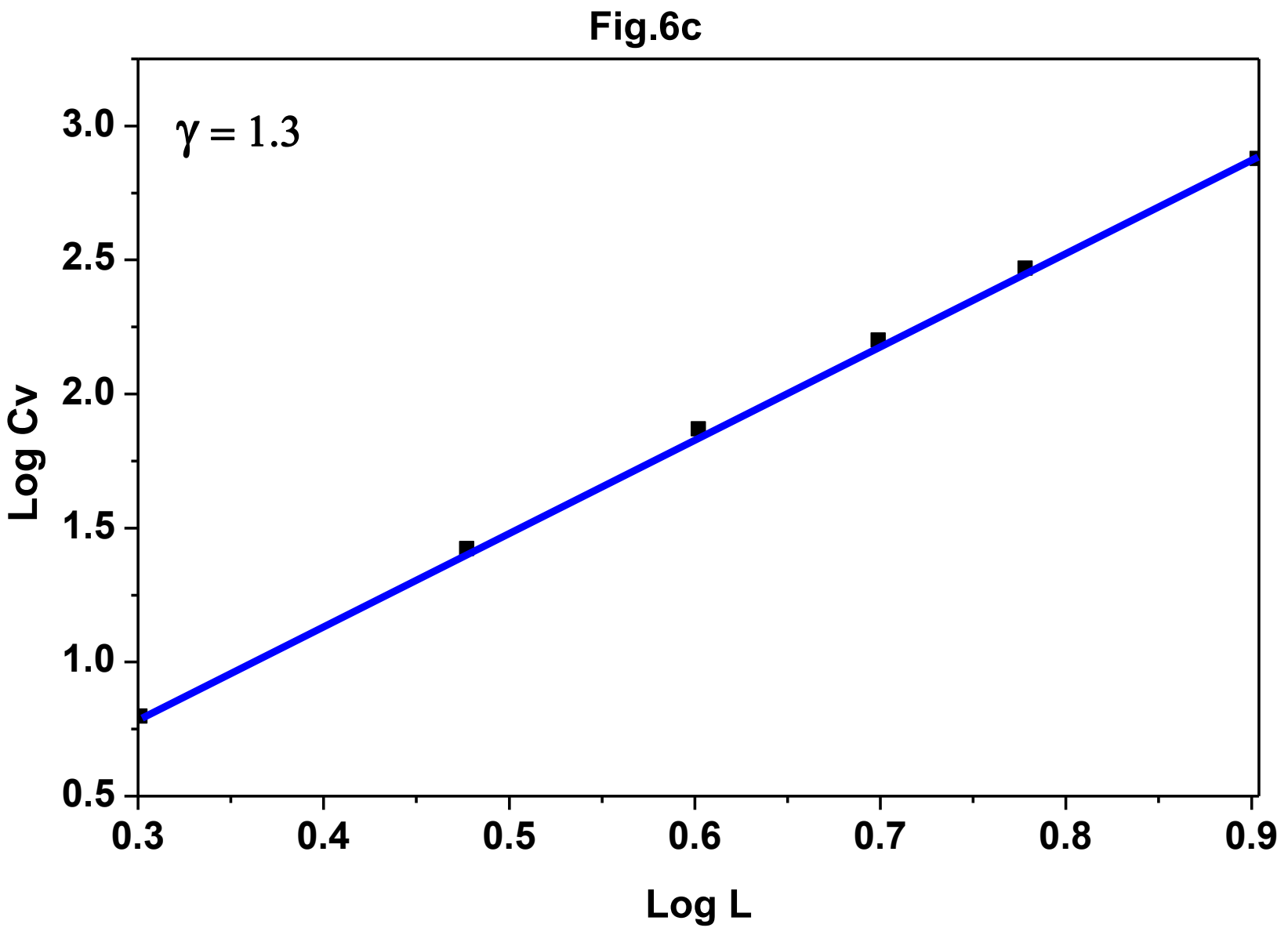

***Fig.6: (a) Log(m) vs. Log(T-Tc); (b) Log(χ) vs. Log(L); (c) Log(Cv) vs. Log(L). The slope in (a) gives the critical exponent α. In (b) the slope corresponds to the critical exponent β. In (c) the slope represents the critical exponent γ.***

| **Critical exponent** | *α* | β | γ |
|---|---|---|---|
| Present study | 0.25 | 0.27 | 1.30 |
| 3D Ising model [46] | 0.20 | 0.30 | 1.20 |

***Table 2. Critical exponent of the studied system GdN.***

## IV. Conclusion

In this paper, we have studied the magnetic properties of the rare earth based compound GdN, using Monte Carlo Simulations (MCS). Firstly, we have analyzed the ground state phase diagrams in the absence of any temperature. At non null temperature values, we discussed the results of the magnetizations, the specific heat and the susceptibilities when varying the different physical parameters. The Monte Carlo simulations are performed in order to investigate the relation between the magnetizations, the susceptibilities, the exchange coupling interactions and the crystal field. To conclude this study, we investigated the hysteresis loops of this system for fixed values of the physical parameters. Finally, we have investigated the critical exponents of the Compound GdN comparing them with those existing in the literature Ref. [46].